\input{aipcheck}

\documentclass[
    ,final            % use final for the camera ready runs
  ]
  {aipproc}

\layoutstyle{8x11double}

\renewcommand\XFMtitleblock{%
  \XFMtitle
  \let\XFMoldpar\par
  \def\par{\XFMoldpar\def\par{\space
             (on behalf of the H.E.S.S.~Collaboration)\XFMoldpar}}%
   \XFMauthors
   \let\par\XFMoldpar
   \XFMaddresses
   \XFMabstract
   \vspace{5pt}%
   \XFMkeywords
   \XFMclassification
 }

\newcommand{\hess}{H.E.S.S.}

\newcommand{\xmm}{{\it XMM-Newton}}
\newcommand{\chandra}{{\it Chandra}}
\newcommand{\suz}{{\it Suzaku}}

\newcommand{\un}[1]{~\hspace{-1pt}\ensuremath{\mathrm{#1}}}

\newcommand{\fvw}{HESS~J1503-582}

\newcommand{\be}{\begin{equation}}
\newcommand{\ee}{\end{equation}}
\newcommand{\ben}{\begin{eqnarray}}
\newcommand{\een}{\end{eqnarray}}
\newcommand{\bc}{\begin{center}}
\newcommand{\ec}{\end{center}}

\def\d{$^\circ$}
\def\m{$^\prime$}
\def\s{$^{\prime\prime}$}
\def\hh{$^{\mathrm h}$}
\def\mm{$^{\mathrm m}$}
\def\ss{$^{\mathrm s}$}

\def\cm3{cm$^{-3}$~}

\def\eg{{\it e.g.~}}
\def\etal{et~al.~}
\def\ie{{\em i.e.~}}

\newcommand\apj{{ApJ}}%
\newcommand\apjs{{ApJS}}%
\newcommand\aap{{A\&A}}%
\begin{document}

\title{On the nature of HESS~J1503-582 revealed by
       the H.E.S.S. experiment: Coincidence with a FVW?}



\author{M. Renaud}{
  address={Max-Planck-Institut f\"ur Kernphysik, Postfach 103980,
69029 Heidelberg, Germany}, email={mrenaud@mpi-hd.mpg.de}}

\author{P. Goret}{
  address={CEA/DSM/IRFU/SAp, L'Orme des Merisiers, 91191
  Gif-sur-Yvette, France}}

\author{and R.C.G. Chaves}{
  address={Max-Planck-Institut f\"ur Kernphysik, Postfach 103980,
69029 Heidelberg, Germany}}

\begin{abstract}

The \hess~survey of the inner Galaxy in the very-high-energy (VHE;
E $>$ 100 GeV) gamma-ray domain has led to the discovery of many
extended sources, some of which do not appear to be associated
with any obvious counterpart at traditional wavelengths (radio,
infrared and X-ray). In this contribution, preliminary
\hess~results on one of these so-called "dark" sources, namely
\fvw, are presented. After introducing the properties of this
source candidate, results of the search for counterparts in
several astronomical windows are shown. Finally, its possible
association with a "Forbidden-Velocity-Wing", a characteristic
21\un{cm} HI line structure that appears as a deviation from the
canonical Galactic rotation curve, is discussed.

\end{abstract}

\maketitle

%%%%%%%%%%%%%%%%%%%%%%%%%%%%%%%%%%%%%%%%%%%%
%% MAINMATTER
%%%%%%%%%%%%%%%%%%%%%%%%%%%%%%%%%%%%%%%%%%%%

\section{Introduction}
\label{s:1}

Almost twenty years after the detection of the first TeV
$\gamma$-ray source, the Crab nebula \cite{c:weekes89}, current
Imaging Atmospheric Cherenkov Telescopes (IACTs) have opened a new
astronomical window with $\sim$ 70 sources detected so far
\cite{c:hinton08}. More than 50 of these new VHE $\gamma$-ray
emitters are of Galactic origin and their detection is largely a
consequence of the survey conducted by \hess, which now covers the
entire inner Galaxy (see \cite{c:chaves08_1}, these proceedings).
A significant fraction of these sources do not appear to be
associated with objects which are known as potential sources of
VHE $\gamma$-rays, such as Supernova Remnants (SNRs) and Pulsar
Wind Nebulae (PWNe). This is likely due to the difficulty of
identifying extended (\ie on the order of tens of arcmins) sources
with no clear sub-structure. Although current IACTs have reached
unprecedented sensitivities and angular resolutions, the
morphology of most of these faint sources can not be characterised
precisely. Moreover, instruments in other domains (radio,
infrared, X-ray) usually feature angular resolutions at the
arcsecond / sub-arcminute scales, often coupled with relatively
small fields of view compared to \hess, which prevent one from
revealing large-scale structures. Generally speaking, every
catalog of potential TeV sources is known to be biased and
incomplete, as exemplified by that of Galactic SNRs
\cite{c:green06}. Therefore, some sources may show up in the VHE
channel while being hardly detectable in other observational
windows. In this contribution, \hess~observations and data
analysis on one of these so-called "dark" sources, namely \fvw,
are introduced. Its properties together with the results from the
search for traditional counterparts such as SNRs, energetic
pulsars and PWNe, star-forming complexes, HII regions and
Wolf-Rayet (WR) stars, are then presented. Finally, its possible
association with a "Forbidden-Velocity-Wing" (hereafter, FVW), a
characteristic 21\un{cm} HI line structure lying beyond the
canonical Galactic rotation curve, is discussed. All results
presented here are preliminary; further studies of this source
candidate are in progress.

\section{\hess~observations \& analysis}
\label{s:2}

\begin{figure}[!htb]

  \includegraphics[height=.22\textheight]{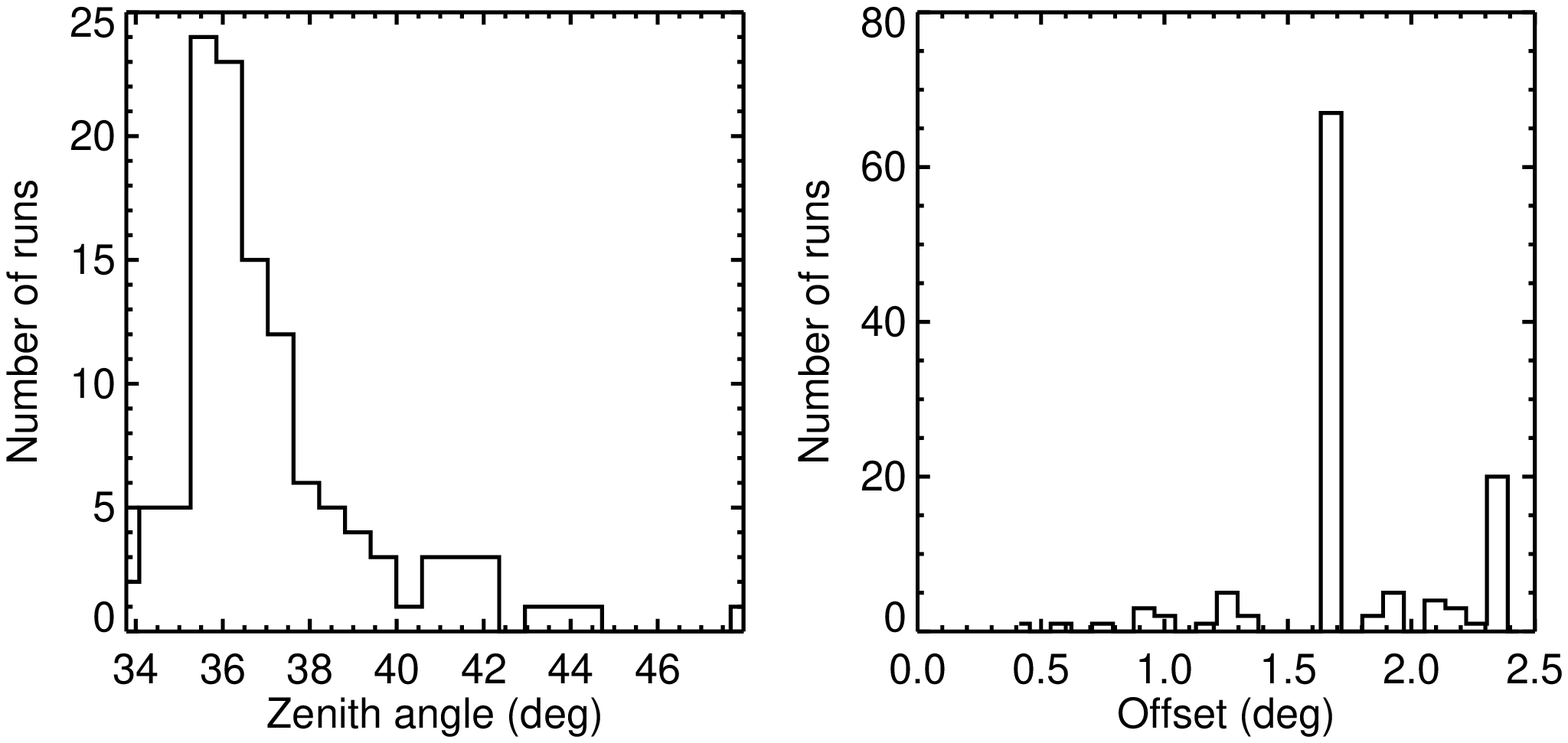}
  \includegraphics[height=.23\textheight]{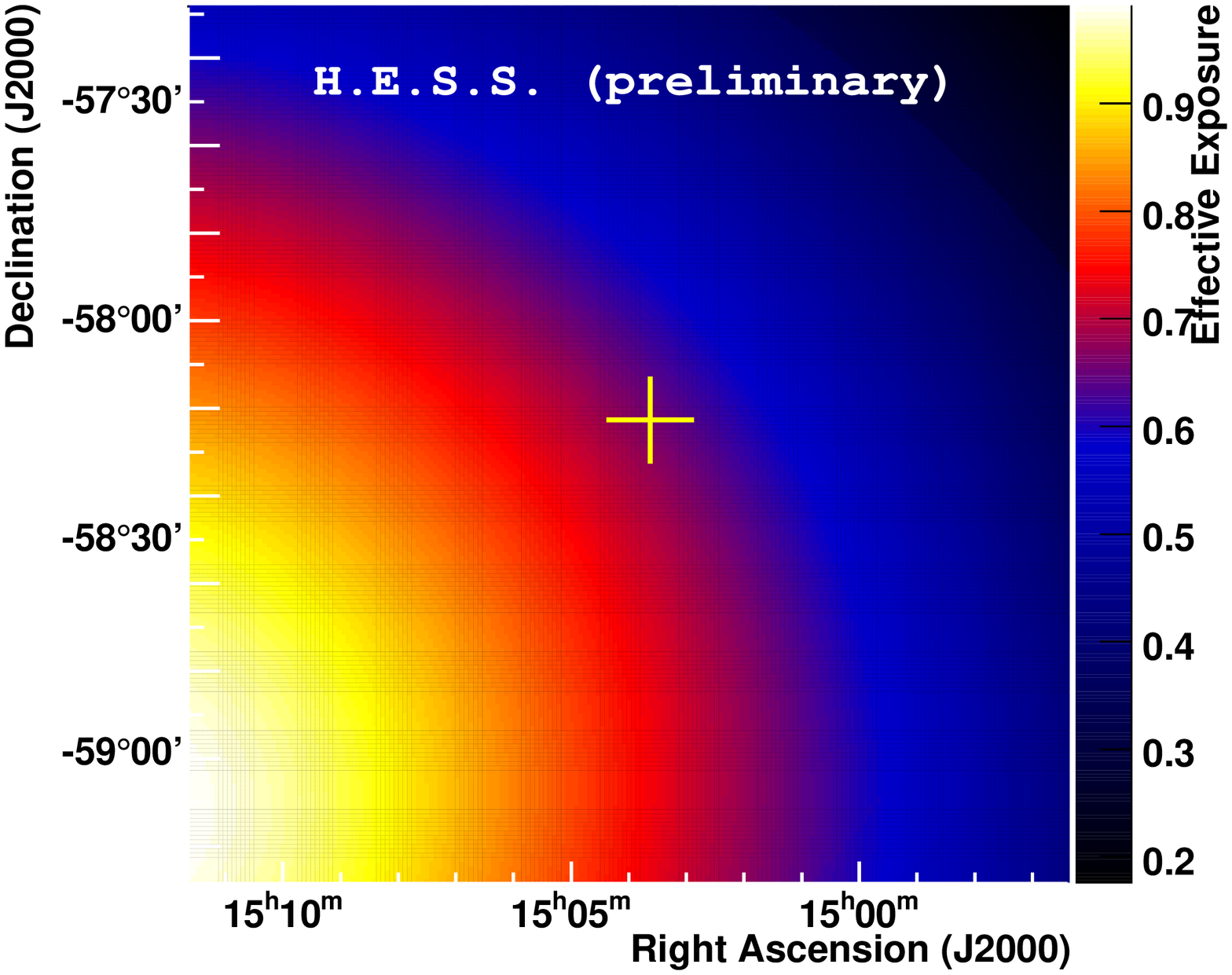}

  \caption{{\em Left:} Distributions of zenith angles and pointing offsets relative to \fvw~of
           selected runs. {\em Right:} Map of the effective exposure (normalized) showing a
           significant exposure gradient in the direction of MSH~15-52, which is $\sim$ 1.5\d~away
           from \fvw. In order to mitigate any potential systematic effects, an additional cut
           was applied, such that only runs with offsets less than 2\d~were selected. The zenith
           angles of the observations range from 35\d~to 45\d, leading to a typical energy
           threshold of about 800\un{GeV}.}
  \label{f:fig1}

\end{figure}

\hess~(High Energy Stereoscopic System) comprises four identical
12\un{m} diameter IACTs located in Namibia. Sensitive to
$\gamma$-rays above $\sim$ 100\un{GeV}, the \hess~array commonly
achieves an angular resolution of about 0.1\d~and an energy
resolution of about 15\%. The region of interest was first
targeted as part of the observational programme on the VHE PWN of
MSH~15-52 in 2004 \cite{c:hess05}. The region was later observed
regularly until 2007 as part of the extended \hess~Galactic Plane
Survey. The data set (see Figure \ref{f:fig1}) was first
investigated using the standard survey analysis (an on-source
region with a radius of $\theta_{cut}$ = 0.22\d, a ring background
region with a radius of 0.8\d~and hard cuts, which include a
minimum requirement of 200 photo-electrons per shower image for
$\gamma$-ray selection) as described in \cite{c:hess06}. An
extended, 5-$\sigma$ excess at l $\sim$ 319.7\d~and b $\sim$
0.3\d~was discovered, namely \fvw.

\begin{figure}[!htb]

  \includegraphics[height=.28\textheight]{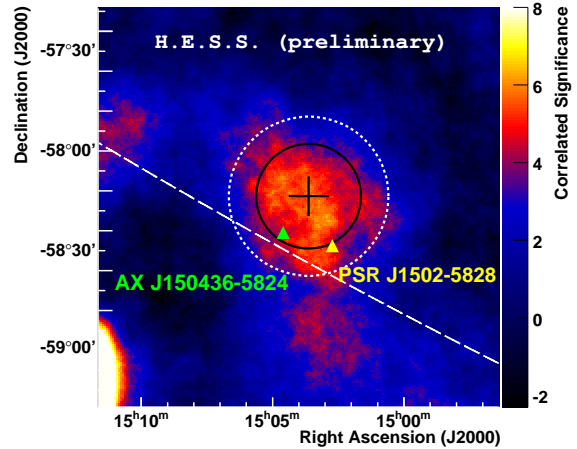}

  \caption{\hess~image of the correlated significance (R~=~0.3\d) centered on \fvw.
           Its peak significance reaches 6 $\sigma$ (pre-trials). The black cross and circle
           denote the uncertainty of the source centroid ($\sim$ 0.1\d) and its intrinsic rms
           size of 0.26\d, respectively, after fitting with a 2D Gaussian. The outer, dotted
           white circle represents the region of spectral extraction. The Galactic Plane is shown
           as the dashed line. The two objects AX~J1504.6-5824 and PSR~J1502-5828 are discussed
           in the next section. The bright source in the lower left corner is MSH~15-52.}
  \label{f:fig2}

\end{figure}

\begin{figure}[!htb]

  \includegraphics[height=.27\textheight]{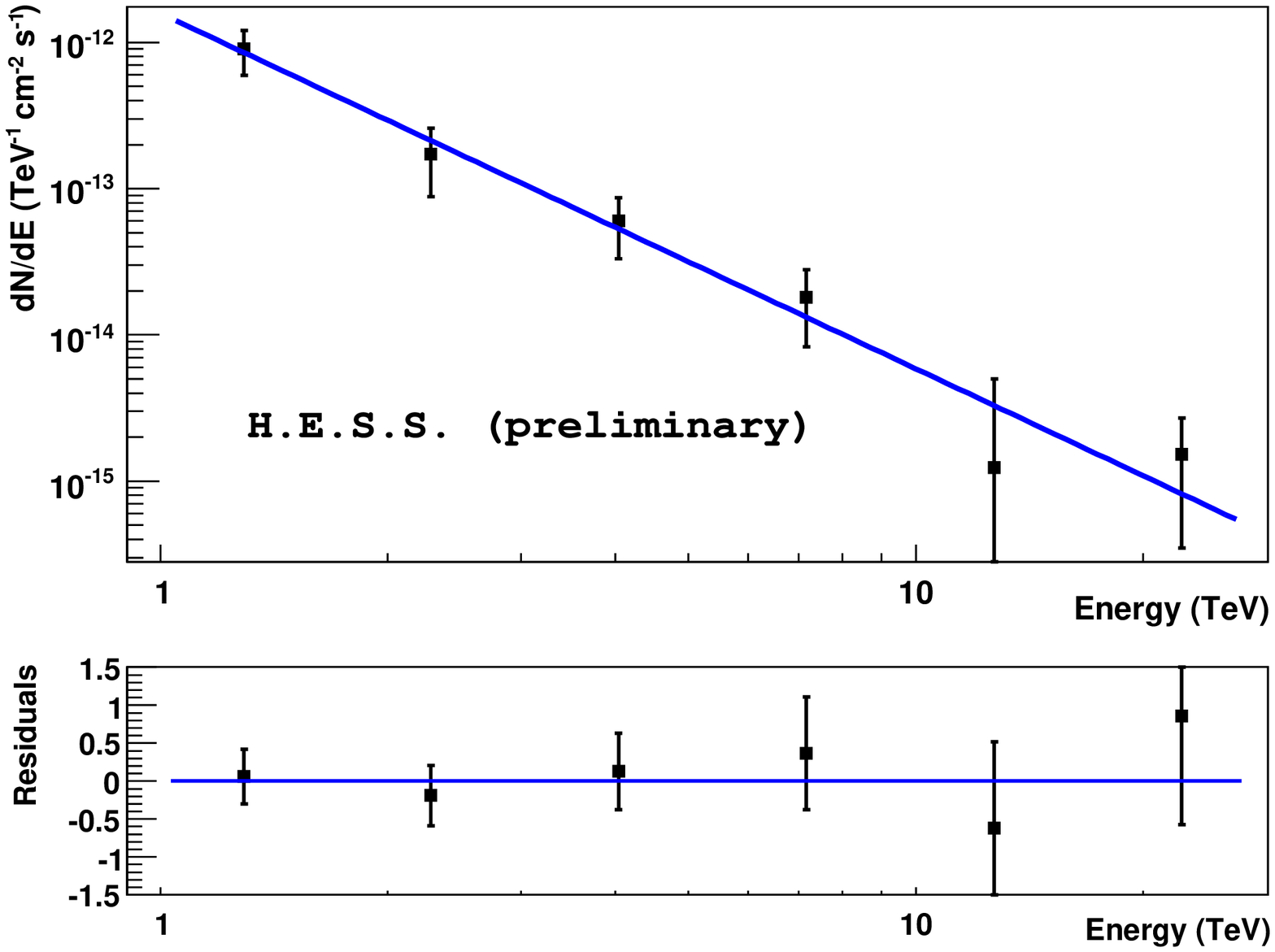}

  \caption{Differential energy spectrum of \fvw~between 1.3 and 22\un{TeV}. The coordinates (J2000)
           of the extraction region are centered on the fit value from Figure \ref{f:fig2}:
           R.A. = 15\hh03\mm38\ss~and Dec. = -58\d13\m45\s~with a radius = 0.4\d. The data points are
           fit with a power law with photon index  $\Gamma$ = 2.4 $\pm$ 0.4$_{stat}$ $\pm$ 0.2$_{syst}$
           and a normalization  at 1\un{TeV} of (1.6 $\pm$ 0.6$_{stat}$) $\times$ 10$^{-12}$ cm$^{-2}$
           s$^{-1}$ TeV$^{-1}$. The integrated flux above 1\un{TeV}, of about 6 $\times$ 10$^{-12}$ erg
           cm$^{-2}$ s$^{-1}$, corresponds to roughly 6 \% of that of the Crab nebula. Also shown
           are the residuals in the bottom panel.}
  \label{f:fig3}

\end{figure}

After selecting only runs with four telescopes which pass the
usual quality criteria (in order to remove data affected by
unstable weather conditions or hardware issues), the data set has
an acceptance-corrected live-time of about 24 h at the position of
\fvw. Two independent data analyses, namely the Hillas
\cite{c:hofmann99} and Model 2D \cite{c:denaurois03} methods, were
used to generate sky maps and spectra. Both analyses give
consistent results. In the following, only those obtained with the
Hillas method are shown. Sky maps (Figure \ref{f:fig2}) were
produced using the Ring Background method for background
subtraction, while spectra (Figure \ref{f:fig3}) were generated
using the Reflected Region Background method \cite{c:berge07}.

\begin{figure}[!htb]

  \includegraphics[height=.24\textheight]{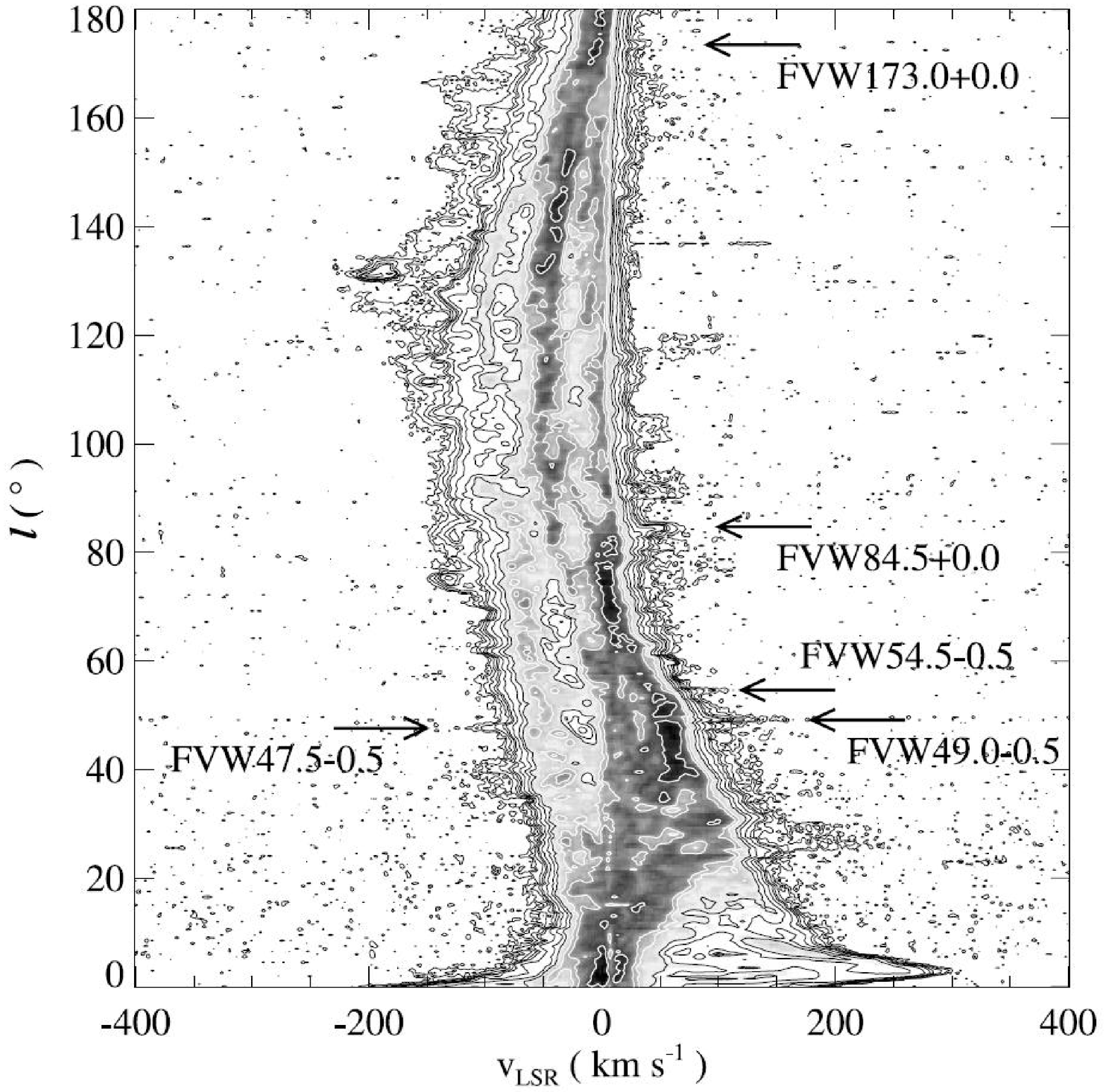}
  \includegraphics[height=.23\textheight]{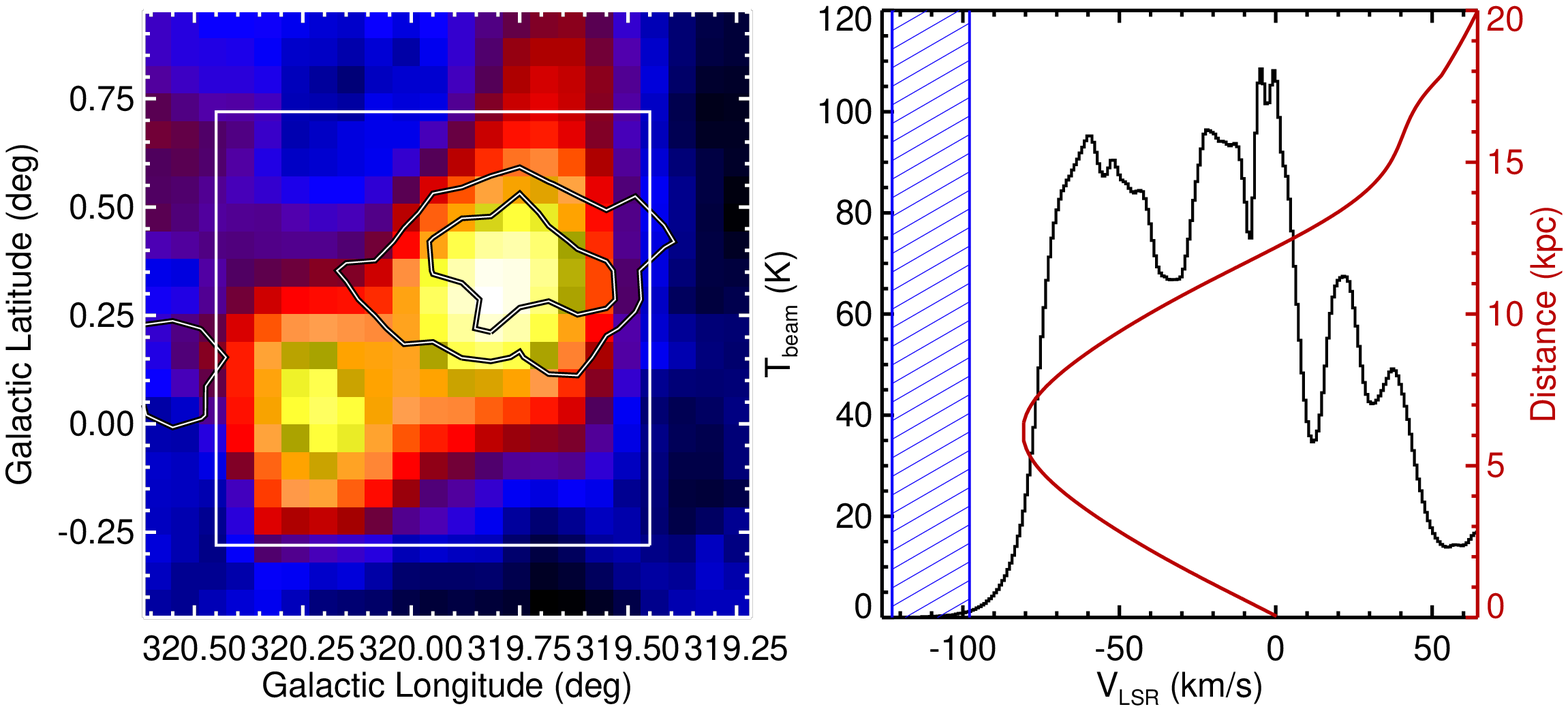}

  \caption{{\em Left:} LDS (l-v) diagram of the first and second quadrants (b = -0.5\d) showing
           some of the FVW structures extending beyond the Galactic Plane (from \cite{c:kk07}). {\em Middle:}
           SGPS (Parkes) velocity-integrated (between -123 and -98 km s$^{-1}$) image of the HI line
           emission centered on FVW~319.8+0.3. \hess~significance contours are shown in black at
           4 and 5 $\sigma$ levels. {\em Right:} Velocity profile of HI intensity integrated over
           the white square shown in the middle panel. The red curve represents the canonical Galactic
           rotation curve according to \cite{c:fich89} at the position of FVW~319.8+0.3. The velocity
           range of the image in the middle is shown by the blue dashed region.}
  \label{f:fig4}

\end{figure}

\section{Multi-wavelength data}
\label{s:3}

\subsection{Search for traditional counterparts}
\label{s:31}

While searching multi-wavelength catalogues (in radio, infrared
and X-ray) within 0.3\d~around \fvw~through the Simbad database,
two interesting sources were found at first glance:

\begin{itemize}

\item AX~J1504.6-5824 \cite{c:sugizaki01}: Catalogued as a
Cataclysmic Variable, given the high column density N(H) = 1.29
$\times$ 10$^{22}$ cm$^{-2}$ and a hard photon index $\Gamma$ =
1.44 in the 2--10\un{keV} band. However, the lack of additional
X-ray data from \xmm, \chandra~or \suz~prevents one from drawing
firm conclusions about its nature.

\item PSR~J1502-5828 \cite{c:atnf08}: An old pulsar ($\tau_{c}$ =
0.29\un{Myr}) with a very low spin-down flux of 3.3 $\times$
10$^{31}$ (d / 12 kpc)$^{-2}$ erg s$^{-1}$ kpc$^{-2}$. This value
correponds to roughly half of the measured integrated flux above
1\un{TeV} of \fvw, which makes the association unlikely; other
$\gamma$-ray emitting PWNe usually exhibit a flux of 10$^{-3}$ -
10$^{-2}$ of the spin-down power \cite{c:carrigan07}.

\end{itemize}

No counterpart was found in catalogues of potential TeV emitters
such as Galactic SNRs \cite{c:green06}, HII regions
\cite{c:paladini03}, star-forming complexes \cite{c:russeil03} and
WR stars \cite{c:vdh01}. The radio archival public images from Molonglo
at 843\un{MHz}, ATCA at 1.4\un{GHz}, Parkes at 2.4\un{GHz}, and
Parkes-MIT-NRAO (PMN) at 4.85\un{GHz}, as well as infrared images
from MSX (at 8.28, 12.13 and 14.65 $\mu$m \footnote{see
\url{http://irsa.ipac.caltech.edu/data/MSX/}}) and Spitzer/GLIMPSE
(at 8 $\mu$m \footnote{see \url{http://irsa.ipac.caltech.edu/data/SPITZER/GLIMPSE/}})
were also inspected to search for hints of diffuse emission
coincident with \fvw, but no such emission was found.

\subsection{Coincidence with a FVW?}
\label{s:32}

The search for traditional counterparts, as discussed above, did
not reveal any likely candidates. However, Kang \& Koo
\cite{c:kk07} have published a catalog of 87 extended and faint
radio structures detected through the 21 cm HI line in the SGPS
and Leiden/Dwingeloo Survey (LDS) data. These structures,
so-called Forbidden-Velocity-Wings (FVWs), appear as wings of line
emission at velocities forbidden by the canonical Galactic
rotation curve in limited spatial regions over velocity extents of
more than $\sim$ 20 km s$^{-1}$ (see Figure \ref{f:fig4}, left).
Among them, FVW~319.8+0.3, marked with the highest detection rank
by Kang \& Koo \cite{c:kk07}, is spatially coincident with \fvw,
as shown in Figure \ref{f:fig4} (middle). This FVW appears in the
HI line image integrated between~-123 and -98 km s$^{-1}$, two
velocities which are not permitted by the canonical Galactic
rotation curve \cite{c:fich89} along this line of sight (Figure
\ref{f:fig4}, right).

This FVW does not coincide with any known objects that could be
responsible for its large velocity, \eg SNRs, nearby galaxies, or
high-velocity clouds. Most of the detected FVWs are located off
the Galactic Plane, and their atypical latitude distribution lead
\cite{c:kk07} to discuss possible origins. For instance, previously
unknown, old SNRs in the radiative phase could be the most likely
candidates, as in the case of the discovery of the SNR associated
with FVW 190.2+1.1 \cite{c:kks06}. Thus, the Southern Galactic
Plane Survey (SGPS) HI data \cite{c:sgps05} have been inspected to
search for any shell-type diffuse emission in the velocity range
of FVW~319.8+0.3, since ATCA features a better angular
resolution ($\sim$ 2\m) than Parkes ($\sim$ 15\m). No shell-type
structure was found, but the low ATCA sensitivity of $\sim$
1.6\un{K} renders the identification of faint and extended
emission difficult.

Such old, isolated SNRs (\ie those with an age greater than the
onset of the radiative phase $\sim$ 3.6 $\times$ 10$^{4}$ y
\cite{c:cioffi88}) are not expected to accelerate multi-TeV
particles any longer, mainly because of the very low shock speed
\cite{c:pz05}. On the other hand, the joint activity of stellar
winds and SN explosions from massive stars in nearby ($<$
4\un{kpc}) and powerful ($>$ 10$^{37}$ erg s$^{-1}$) OB
associations could produce fast-moving neutral hydrogen gas
detectable at the sensitivity level of the current HI surveys
\cite{c:kk07}. Recently, \hess~detected extended VHE $\gamma$-ray
emission in the direction of the young stellar cluster
Westerlund~2, hosting the massive WR binary WR~20a and located in
the HII complex RCW~49 \cite{c:hess07}. This source is currently
the only clear exemple of an association between VHE $\gamma$-ray
emission and a young stellar cluster~/ wind-blown bubble
(\cite{c:chaves08_2} discuss the status of HESS~J1848-018 in this
regard). However, it would not be unlikely if, among all of the
known "dark" sources, some of them actually belong to this class
of VHE $\gamma$-ray emitters. Deeper investigations of infrared
data and follow-up observations of \fvw~will then help constrain
such scenario.

\section{Conclusion}
\label{s:4}

The \hess~Cherenkov telescope array has proven itself to be the
most efficient in revealing new faint and extended VHE
$\gamma$-ray sources. Among them, the source candidate \fvw~does
not appear to have any obvious counterpart at traditional
wavelengths. With an observing strategy designed to reduce the
gradient in the current exposure map, more data will soon be taken
in order to confirm this detection. Follow-up X-ray observations
with \xmm, \chandra~and \suz~are also needed in order to constrain
the nature of the ASCA hard X-ray point-like source close to \fvw.
The association with a FVW, if confirmed, would represent the
first source of VHE $\gamma$-rays coincident with such an HI
structure. These structures could be the result of the combined
activity of stellar winds and supernova explosions that are
detectable only through this channel.

\begin{theacknowledgments}

The support of the Namibian authorities and of the University of
Namibia in facilitating the construction and operation of \hess~is
gratefully acknowledged, as is the support by the German Ministry
for Education and Research (BMBF), the Max Planck Society, the
French Ministry for Research, the CNRS-IN2P3 and the Astroparticle
Interdisciplinary Programme of the CNRS, the U.K. Science and
Technology Facilities Council (STFC), the IPNP of the Charles
University, the Polish Ministry of Science and Higher Education,
the South African Department of Science and Technology and
National Research Foundation, and by the University of Namibia. We
appreciate the excellent work of the technical support staff in
Berlin, Durham, Hamburg, Heidelberg, Palaiseau, Paris, Saclay, and
in Namibia in the construction and  operation of the equipment.
This research has made use of the SIMBAD database, operated at
CDS, Strasbourg, France. We would also like to thank B.-C. Koo and
J.-H. Kang for fruitful discussions.

\end{theacknowledgments}

\bibliographystyle{aipproc}   % if natbib is available
%\bibliographystyle{aipprocl} % if natbib is missing

%\bibliography{sample}
%\IfFileExists{\jobname.bbl}{}
% {\typeout{}
%  \typeout{******************************************}
%  \typeout{** Please run "bibtex \jobname" to optain}
%  \typeout{** the bibliography and then re-run LaTeX}
%  \typeout{** twice to fix the references!}
%  \typeout{******************************************}
%  \typeout{}
% }

\end{document}